\documentclass[10pt,twocolumn,letterpaper]{article}

\usepackage{cvpr}
\usepackage{times}
\usepackage{graphicx}
\usepackage{amsmath}
\usepackage{amssymb}
\usepackage{multirow}
\usepackage{color}
\usepackage{array,ragged2e}
\usepackage{booktabs}
\usepackage{paralist}
\usepackage{rotating}

\def\eg{{\em e.g.}}
\def\ie{{\em i.e.}}

\usepackage[colorlinks=true,urlcolor=magenta,citecolor=blue,breaklinks=true,bookmarks=false]{hyperref}

\cvprfinalcopy 

\def\cvprPaperID{2280} 

\ifcvprfinal\fi
\begin{document}

\title{Celeb-DF: A Large-scale Challenging Dataset for DeepFake Forensics}

\author{Yuezun Li$^1$, Xin Yang$^1$, Pu Sun$^2$, Honggang Qi$^2$ and Siwei Lyu$^1$ \\
$^1$ University at Albany, State University of New York, USA \\
$^2$ University of Chinese Academy of Sciences, China}

\maketitle
\thispagestyle{empty}

\begin{abstract}
AI-synthesized face-swapping videos, commonly known as {\em DeepFakes}, is an emerging problem threatening the trustworthiness of online information. The need to develop and evaluate DeepFake detection algorithms calls for large-scale datasets. However, current DeepFake datasets suffer from low visual quality and do not resemble DeepFake videos circulated on the Internet. We present a new large-scale challenging DeepFake video dataset, {\em Celeb-DF}, which contains {$5,639$} high-quality DeepFake videos of celebrities generated using improved synthesis process.  We conduct a comprehensive evaluation of DeepFake detection methods and datasets to demonstrate the escalated level of challenges posed by Celeb-DF. 
\end{abstract}

\begin{figure}[h]
\begin{tabular}{@{\hspace{0em}}l@{\hspace{0em}}l}
   \raisebox{3em}{
   \begin{turn}{90}
   UADFV
   \end{turn}
   }
   &\includegraphics[width=.45\textwidth]{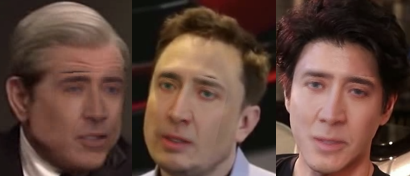} \\
   \raisebox{2em}{
   \begin{turn}{90}
   DF-TIMIT-HQ
   \end{turn}
   }
    &\includegraphics[width=.45\textwidth]{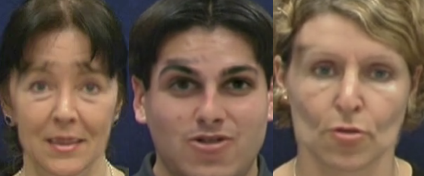} \\
    \raisebox{3em}{
   \begin{turn}{90}
   FF-DF
   \end{turn}
   }
    &\includegraphics[width=.45\textwidth]{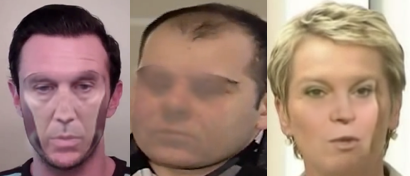} \\
    \raisebox{4em}{
   \begin{turn}{90}
   DFD
   \end{turn}
   }&\includegraphics[width=.45\textwidth]{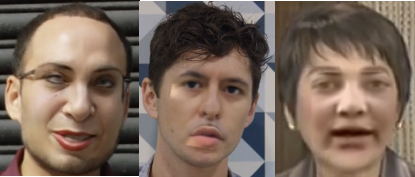} \\
    \raisebox{3em}{
   \begin{turn}{90}
   DFDC
   \end{turn}
   }&\includegraphics[width=.45\textwidth]{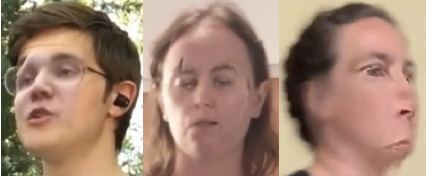}
\end{tabular}
\caption{\em \small Visual artifacts of DeepFake videos in existing datasets. Note some common types of visual artifacts in these video frames, including low-quality synthesized faces (row 1 col 1, row 3 col 2, row 5 col 3), visible splicing boundaries (row 3 col 1, row 4 col 2, row 5 col 2), color mismatch (row 5 col 1), visible parts of the original face (row 1 col 1, row 2 col 1, row 4 col 3), and inconsistent synthesized face orientations (row 3 col 3). This figure is best viewed in color.}
    \label{fig:overview}
    \vspace{-2em}
\end{figure}

\begin{figure*}[t]
    \centering
    \includegraphics[width=\linewidth]{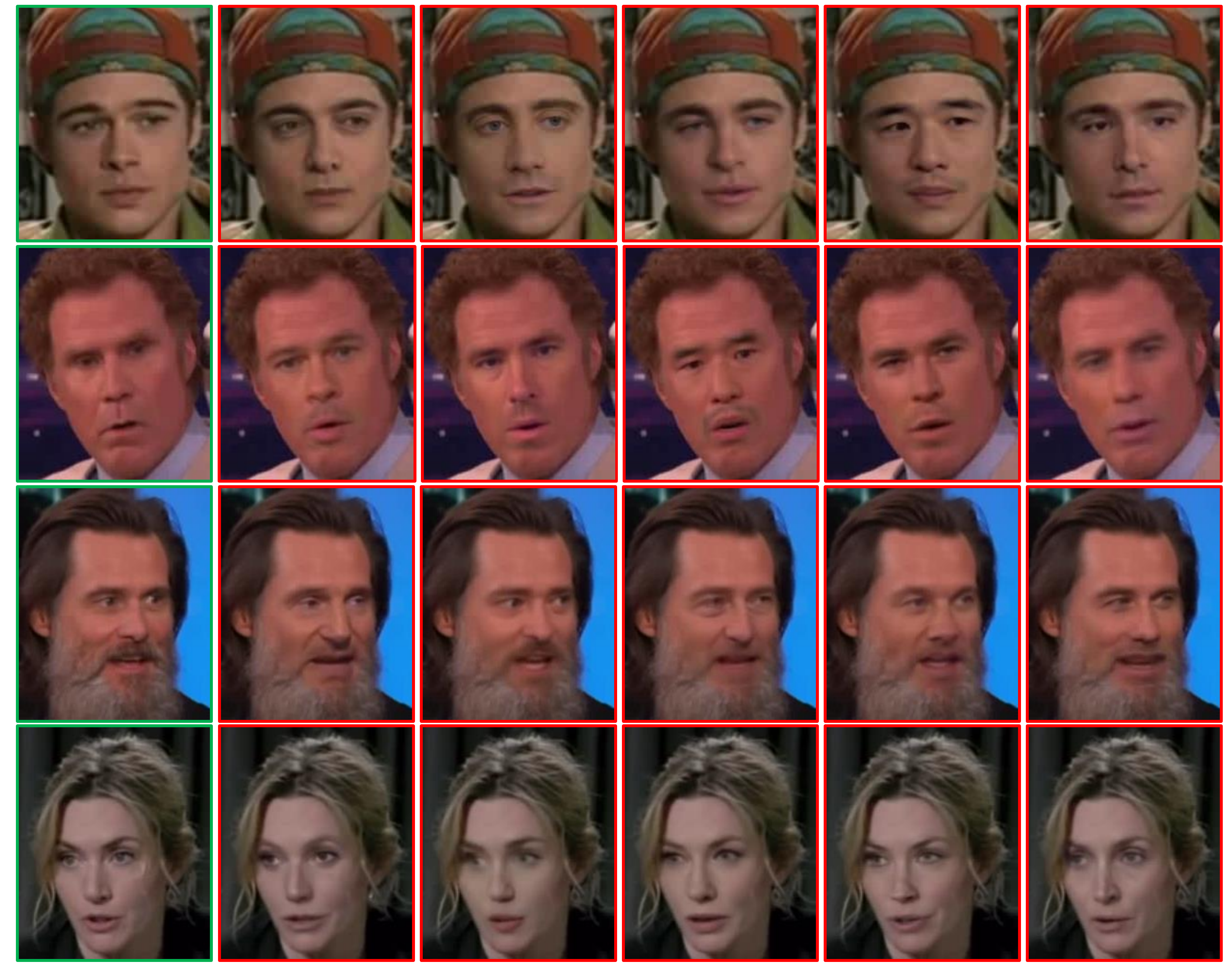}
    \vspace{-0.5cm}
    \caption{\em \small Example frames from the Celeb-DF dataset. Left column is the frame of real videos and right five columns are corresponding DeepFake frames generated using different donor subject.}
    \label{fig:demo}
    \vspace{-1em}
\end{figure*}

\section{Introduction}

A recent twist to the disconcerting problem of online disinformation is falsified videos created by AI technologies, in particular, deep neural networks (DNNs). Although fabrication and manipulation of digital images and videos are not new \cite{farid08}, the use of DNNs has made the process to create convincing fake videos increasingly easier and faster. 

One particular type of DNN-based fake videos, commonly known as {\em DeepFakes}, has recently drawn much attention. In a DeepFake video, the faces of a {\em target} individual are replaced by the faces of a {\em donor} individual synthesized by DNN models, retaining the target's facial expressions and head poses. Since faces are intrinsically associated with identity, well-crafted DeepFakes can create illusions of a person's presence and activities that do not occur in reality, which can lead to serious political, social, financial, and legal consequences \cite{survey_chesney_citron_2018}.

With the escalated concerns over the DeepFakes, there is a surge of interest in developing DeepFakes detection methods recently  \cite{afchar2018mesonet,guera2018deepfake,li2018ictu,yang2018exposing,matern2019exploiting,li2019exposing,sabir2019recurrent,roessler2019faceforensics++,nguyen2019capsule,nguyen2019multi,nguyen2019capsulev2}, with an upcoming dedicated global {\em DeepFake Detection Challenge}\footnote{\url{https://deepfakedetectionchallenge.ai}.}. The availability of large-scale datasets of DeepFake videos is an enabling factor in the development of DeepFake detection method. To date, we have the UADFV dataset \cite{yang2018exposing}, the DeepFake-TIMIT dataset (DF-TIMIT) \cite{korshunov2018deepfakes}, the FaceForenscics++ dataset (FF-DF) \cite{roessler2019faceforensics++}\footnote{ FaceForensics++ contains other types of fake videos. We consider only the DeepFake videos.}, the Google DeepFake detection dataset (DFD) \cite{DDD_GoogleJigSaw2019}, and the FaceBook DeepFake detection challenge (DFDC) dataset \cite{dolhansky2019deepfake}. 

However, a closer look at the DeepFake videos in existing datasets reveals stark contrasts in visual quality to the actual DeepFake videos circulated on the Internet. Several common visual artifacts that can be found in these datasets are highlighted in Fig.\ref{fig:overview}, including low-quality synthesized faces, visible splicing boundaries, color mismatch, visible parts of the original face, and inconsistent synthesized face orientations. These artifacts are likely the result of imperfect steps of the synthesis method and the lack of curating of the synthesized videos before included in the datasets. Moreover, DeepFake videos with such low visual qualities can hardly be convincing, and are unlikely to have real impact. Correspondingly, high detection performance on these dataset may not bear strong relevance when the detection methods are deployed {\em in the wild}.
  
In this work, we present a new large-scale and challenging DeepFake video dataset, {\em Celeb-DF}\footnote{\url{http://www.cs.albany.edu/~lsw/celeb-deepfakeforensics.html}.}, for the development and evaluation of DeepFake detection algorithms. There are in total {$5,639$} DeepFake videos, corresponding more than $2$ million frames, in the Celeb-DF dataset. The real source videos are based on publicly available {\tt YouTube} video clips of {$59$} celebrities of diverse genders, ages, and ethic groups. The DeepFake videos are generated using an improved DeepFake synthesis method.  As a result, the overall visual quality of the synthesized DeepFake videos in Celeb-DF is greatly improved when compared to existing datasets, with significantly fewer notable visual artifacts, see Fig.\ref{fig:demo}. Based on the Celeb-DF dataset and other existing datasets, we conduct an evaluation of current DeepFake detection methods. This is the most comprehensive performance evaluation of DeepFake detection methods to date. The results show that Celeb-DF is challenging to most of the existing detection methods, even though many DeepFake detection methods are shown to achieve high, sometimes near perfect, accuracy on previous datasets. 


\section{Backgrounds}
\subsection{DeepFake Video Generation}

Although in recent years there have been many sophisticated algorithms for generating realistic synthetic face videos \cite{bitouk2008face,dale2011video,suwajanakorn2015makes,Thies_2016_CVPR,korshunova2017fast,suwajanakorn2017synthesizing,pham2018generative,karras2018progressive,kim2018DeepVideo,chan2019everybody,karras2019style,thies2019deferred}, most of these have not been in mainstream as open-source software tools that anyone can use. It is a much simpler method based on the work of neural image style transfer \cite{liu2017unsupervised} that becomes the {\em tool of choice} to create DeepFake videos in scale, with several independent open-source implementations, \eg, {\tt FakeApp} \cite{fakeapp}, {\tt DFaker} \cite{DFaker}, {\tt faceswap-GAN} \cite{faceswap-gan}, {\tt faceswap} \cite{faceswap}, and {\tt DeepFaceLab} \cite{DeepFaceLab}. We refer to this method as the {\em basic DeepFake maker}, and it is underneath many DeepFake videos circulated on the Internet or in the existing datasets. 

\begin{figure*}[t]
\centering
\includegraphics[width=.8\linewidth]{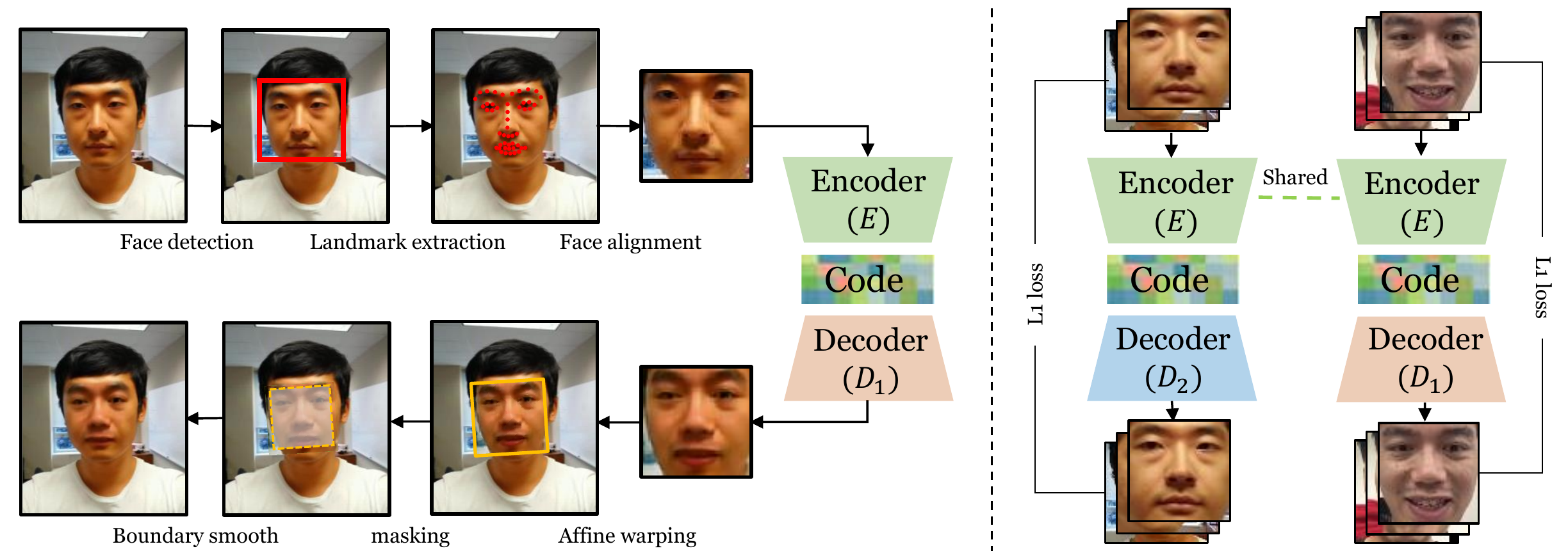}
\vspace{-.3em}
\caption{\small \em Synthesis (left) and training (right) of the basic DeepFake maker algorithm. See texts for more details.}
\label{fig:pipeline}
\vspace{-1.5em}
\end{figure*}

The overall pipeline of the basic DeepFake maker is shown in Fig.\ref{fig:pipeline} (left). From an input video, faces of the target are detected, from which facial landmarks are further extracted. The landmarks are used to align the faces to a standard configuration \cite{kazemi2014one}. The aligned faces are then cropped and fed to an auto-encoder \cite{kingma2013auto} to synthesize faces of the donor with the same facial expressions as the original target's faces. 

The auto-encoder is usually formed by two convoluntional neural networks (CNNs), \ie, the {\em encoder} and the {\em decoder}. The encoder $E$ converts the input target's face to a vector known as the {\em code}. To ensure the encoder capture identity-independent attributes such as facial expressions, there is one single encoder regardless the identities of the subjects. On the other hand, each identity has a dedicated decoder $D_i$, which generates a face of the corresponding subject from the code. The encoder and decoder are trained in tandem using uncorresponded face sets of multiple subjects in an unsupervised manner, Fig.\ref{fig:pipeline} (right). Specifically, an encoder-decoder pair is formed alternatively using $E$ and $D_i$ for input face of each subject, and optimize their parameters to minimize the reconstruction errors ($\ell_1$ difference between the input and reconstructed faces). The parameter update is performed with the back-propagation until convergence. 

The synthesized faces are then warped back to the configuration of the original target's faces and trimmed with a {\em mask} from the facial landmarks. The last step involves smoothing the boundaries between the synthesized regions and the original video frames. The whole process is automatic and runs with little manual intervention. 

\subsection{DeepFake Detection Methods}

Since DeepFakes become a global phenomenon, there has been an increasing interest in DeepFake detection methods. Most of the current DeepFake detection methods use data-driven deep neural networks (DNNs) as backbone. 

Since synthesized faces are spliced into the original video frames, state-of-the-art DNN splicing detection methods, \eg,~\cite{zhou2017two,zhou2018learning,liu2018image,bappy2019hybrid}, can be applied. There have also been algorithms dedicated to the detection of DeepFake videos that fall into three categories.  Methods in the first category are based on inconsistencies exhibited in the {\bf physical/physiological} aspects in the DeepFake videos. The method in work of \cite{li2018ictu} exploits the observation that many DeepFake videos lack reasonable eye blinking due to the use of online portraits as training data, which usually do not have closed eyes for aesthetic reasons. Incoherent head poses in DeepFake videos are utilized in \cite{yang2018exposing} to expose DeepFake videos.  In~\cite{agarwal2019protecting}, the idiosyncratic behavioral patterns of a particular individual are captured by the time series of facial landmarks extracted from real videos are used to spot DeepFake videos. The second category of DeepFake detection algorithms (\eg, \cite{matern2019exploiting,li2019exposing}) use {\bf signal-level} artifacts introduced during the synthesis process such as those described in the Introduction. The third category of DeepFake detection methods (\eg, \cite{afchar2018mesonet,guera2018deepfake,nguyen2019capsule,nguyen2019capsulev2}) are {\bf data-driven}, which directly employ various types of DNNs trained on real and DeepFake videos, not relying on any specific artifact. 

\begin{table}[t]
\small
\centering
  \resizebox{\linewidth}{!}{
  \begin{tabular}{|c|c|c|c|c|c|}
    \hline
    \multirow{2}{*}{Dataset} & \multicolumn{2}{c|}{$\#$ Real} & \multicolumn{2}{c|}{$\#$ DeepFake} & \multirow{2}{*}{Release Date} \\
    \cline{2-5}
    & Video & Frame & Video & Frame & \\
    \hline
    \hline
    UADFV  & 49 & 17.3k & 49 & 17.3k & 2018.11 \\
    \hline
    DF-TIMIT-LQ & \multirow{2}{*}{320$^*$} & \multirow{2}{*}{34.0k} & 320 & 34.0k &  \multirow{2}{*}{2018.12} \\
    DF-TIMIT-HQ & & & 320 & 34.0k & \\
    \hline
    FF-DF  & 1,000 & 509.9k & 1,000 & 509.9k & 2019.01 \\
    \hline
    DFD & 363 & 315.4k & 3,068 & 2,242.7k & 2019.09 \\
    \hline
    DFDC & 1,131 & 488.4k & 4,113 & 1,783.3k & 2019.10 \\
    \hline
    \hline
\bf Celeb-DF & 590 & 225.4k & {\bf 5,639} & 2,116.8k & 2019.11 \\
    \hline
  \end{tabular}
  }
  \caption{\em \small Basic information of various DeepFake video datasets. $*$: the original videos in DF-TIMIT are from Vid-TIMIT dataset.}
  \label{table:stat}
  \vspace{-2em}
\end{table}


\subsection{Existing DeepFake Datasets}
\label{sec:prev-dbs}

DeepFake detection methods require training data and need to be evaluated. As such, there is an increasing need for large-scale DeepFake video datasets. Table \ref{table:stat} lists the current DeepFake datasets.

\noindent{\bf UADFV}: The UADFV dataset \cite{yang2018exposing} contains $49$ real {\tt YouTube} and $49$ DeepFake videos. The DeepFake videos are generated using the DNN model with {\tt FakeAPP} \cite{fakeapp}. 

\noindent{\bf DF-TIMIT}: The DeepFake-TIMIT dataset \cite{korshunov2018deepfakes} includes $640$ DeepFake videos generated with {\tt faceswap-GAN} \cite{faceswap-gan} and based on the Vid-TIMIT dataset \cite{sanderson2009multi}. The videos are divided into two equal-sized subsets: DF-TIMIT-LQ and DF-TIMIT-HQ, with synthesized faces of size $64 \times 64$ and $128 \times 128$ pixels, respectively.

\noindent{\bf FF-DF}: The FaceForensics++ dataset \cite{roessler2019faceforensics++} includes a subset of DeepFakes videos, which has $1,000$ real {\tt YouTube} videos and the same number of synthetic videos generated using {\tt faceswap} \cite{faceswap}.

\noindent{\bf DFD}: The Google/Jigsaw DeepFake detection dataset \cite{DDD_GoogleJigSaw2019} has $3,068$ DeepFake videos generated based on $363$ original videos of $28$ consented individuals of various genders, ages and ethnic groups. The details of the synthesis algorithm are not disclosed, but it is likely to be an improved implementation of the basic DeepFake maker algorithm. 

\noindent{\bf DFDC}: The Facebook DeepFake detection challenge dataset \cite{dolhansky2019deepfake} is part of the DeepFake detection challenge, which has $4,113$ DeepFake videos created based on $1,131$ original videos of $66$ consented individuals of various genders, ages and ethnic groups\footnote{The full set of DFDC has not been released at the time of CVPR submission, and information is based on the first round release in \cite{dolhansky2019deepfake}.}. {This dataset is created using two different synthesis algorithms, but the details of the synthesis algorithm are not disclosed.}

Based on release time and synthesis algorithms, we categorize UADFV, DF-TIMIT, and FF-DF as the {\em first generation} of DeepFake datasets, while DFD, DFDC, and the proposed Celeb-DF datasets are the {\em second generation}. In general, the second generation datasets improve in both quantity and quality over the first generation. 

\section{The Celeb-DF Dataset}

Although the current DeepFake datasets have sufficient number of videos, as discussed in the Introduction and demonstrate in Fig.\ref{fig:overview}, DeepFake videos in these datasets have various visual artifacts that easily distinguish them from the real videos. To provide more relevant data to evaluate and support the future development DeepFake detection methods, we construct the Celeb-DF dataset. A comparison of the Celeb-DF dataset with other existing DeepFake datasets is summarized in Table \ref{table:stat}. 

\subsection{Basic Information}

The Celeb-DF dataset is comprised of {$590$} real videos and {$5,639$} DeepFake videos (corresponding to over two million video frames). The average length of all videos is approximate $13$ seconds with the standard frame rate of $30$ frame-per-second. The real videos are chosen from publicly available {\tt YouTube} videos, corresponding to interviews of {$59$} celebrities with a diverse distribution in their genders, ages, and ethnic groups\footnote{We choose celebrities' faces as they are more familiar to the viewers so that any visual artifacts can be more readily identified. Furthermore, celebrities are anecdotally the main targets of DeepFake videos.}. $56.8\%$ subjects in the real videos are male, and $43.2\%$ are female. $8.5\%$ are of age 60 and above, $30.5\%$ are between 50 - 60, $26.6\%$ are 40s, $28.0\%$ are 30s, and $6.4\%$ are younger than 30. $5.1\%$ are Asians, $6.8\%$ are African Americans and $88.1\%$ are Caucasians. In addition, the real videos exhibit large range of changes in aspects such as the subjects' face sizes (in pixels), orientations, lighting conditions, and backgrounds. The DeepFake videos are generated by swapping faces for each pair of the {$59$} subjects. The final videos are in MPEG4.0 format.

\subsection{Synthesis Method}
\label{sec:alg}

The DeepFake videos in Celeb-DF are generated using an improved DeepFake synthesis algorithm, which is key to the improved visual quality as shown in Fig.\ref{fig:demo}. Specifically, the basic DeepFake maker algorithm is refined in several aspects targeting the following specific visual artifacts observed in existing datasets.

\begin{figure}
\centering
\includegraphics[width=\linewidth]{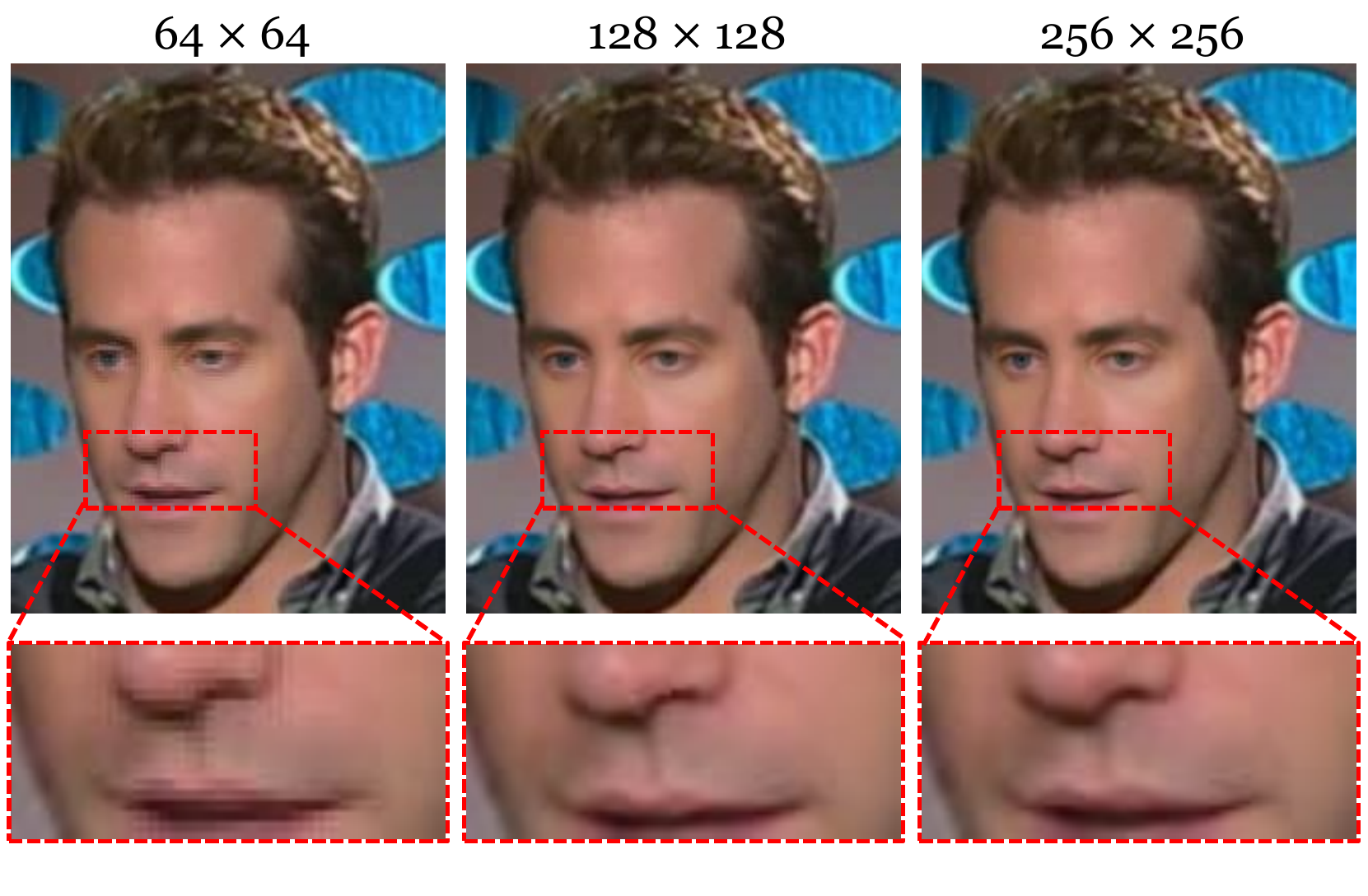}
\vspace{-2.em}
\caption{\small \em Comparison of DeepFake frames with different sizes of the synthesized faces. Note the improved smoothness of the $256 \times 256$ synthesized face, which is used in Celeb-DF. This figure is best viewed in color.}
\label{fig:resolution_compare}
\vspace{-1.5em}
\end{figure}

\smallskip
\noindent{\bf Low resolution of synthesized faces}: The basic DeepFake maker algorithm generate low-resolution faces (typically $64 \times 64$ or $128 \times 128$ pixels). We improve the resolution of the synthesized face to $256 \times 256$ pixels. This is achieved by using encoder and decoder models with more layers and increased dimensions. We determine the structure empirically for a balance between increased training time and better synthesis result. The higher resolution of the synthesized faces are of better visual quality and less affected by resizing and rotation operations in accommodating the input target faces, Fig.\ref{fig:resolution_compare}.

\begin{figure}[h]
\centering
\includegraphics[width=0.85\linewidth]{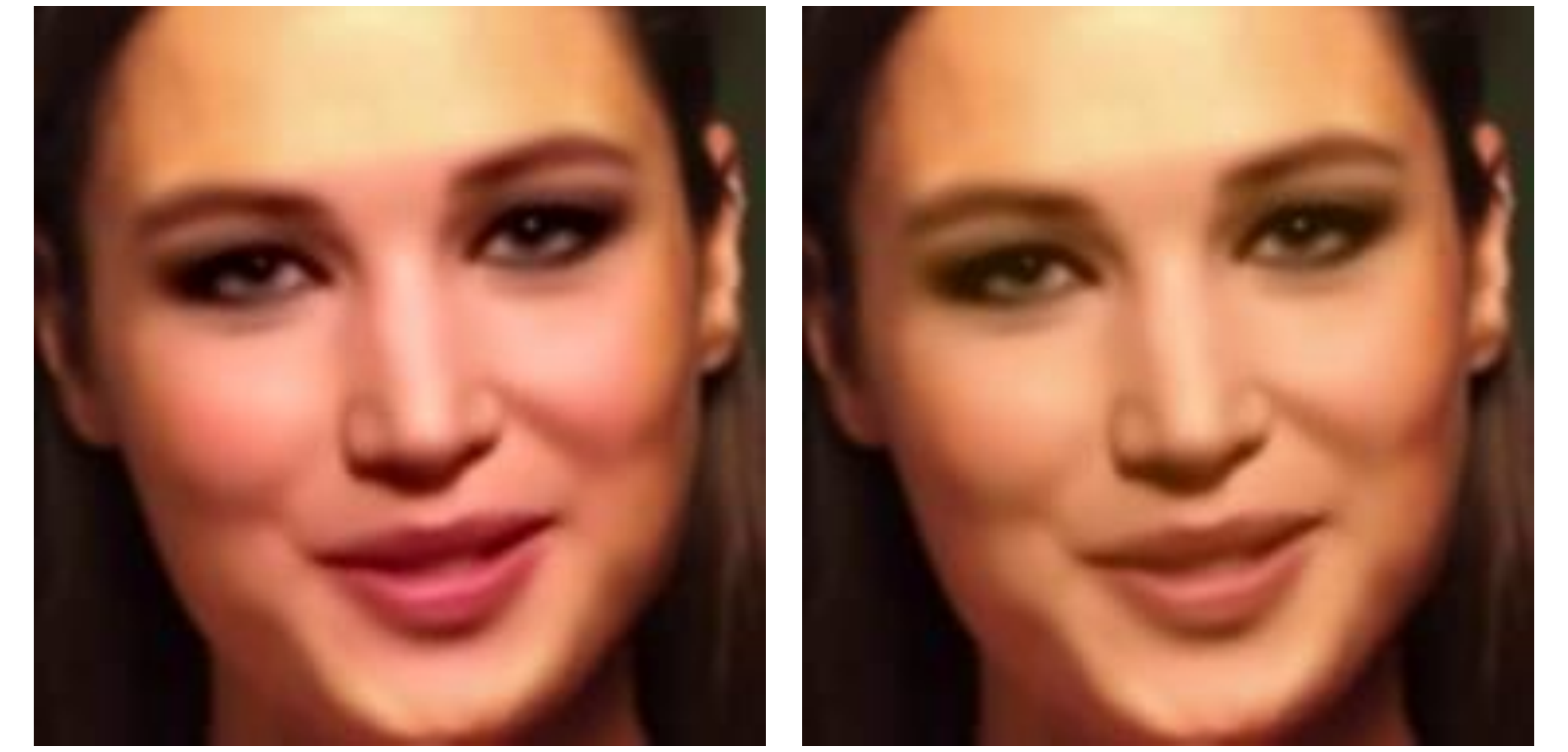}
\caption{\small \em DeepFake frames using synthesized face without (left) and with (right) color correction. Note the reduced color mismatch between the synthesized face region and the other part of the face. Synthesis method with color correction is used to generate Celeb-DF. This figure is best viewed in color.}
\label{fig:color_transfer}
\vspace{-.5em}
\end{figure}

\smallskip
\noindent{\bf Color mismatch}: Color mismatch between the synthesized donor's face with the original target's face in Celeb-DF is significantly reduced by training data augmentation and post processing. Specifically, in each training epoch, we randomly perturb the colors of the training faces, which forces the DNNs to synthesize an image containing the same color pattern with input image. We also apply a color transfer algorithm \cite{reinhard2001color} between the synthesized donor face and the input target face. Fig.\ref{fig:color_transfer} shows an example of synthesized face without (left) and with (right) color correction.
\vspace{-0.3cm}

\begin{figure}
\centering
\includegraphics[width=\linewidth]{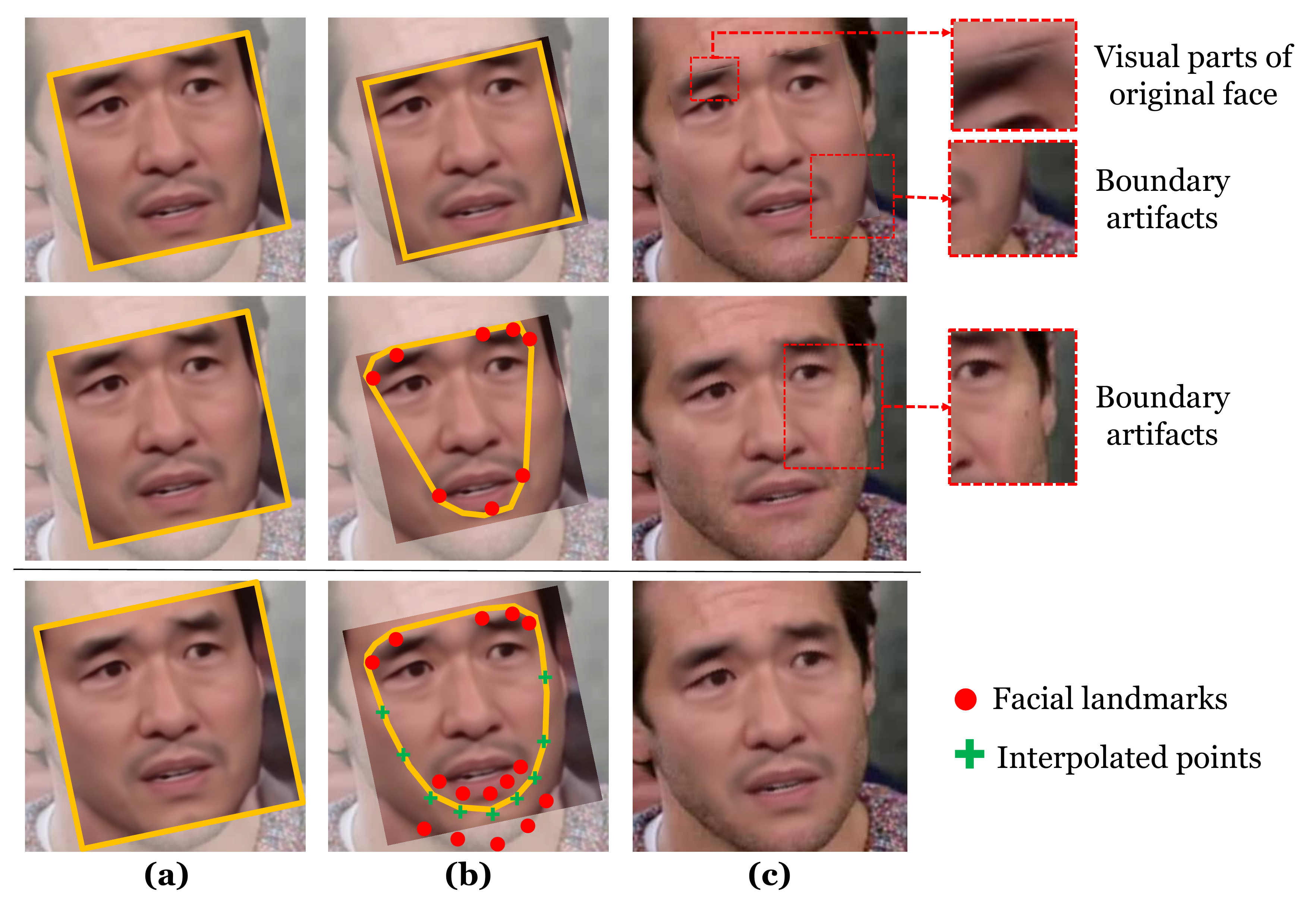}
\vspace{-1.5em}
\caption{\small \em Mask generation in existing datasets (Top two rows) and Celeb-DF (3rd row). (a) warped synthesized face overlaying the target's face. (b) mask generation. (c) final synthesis result.}
\label{fig:mask}
\vspace{-1.5em}
\end{figure}

\smallskip
\noindent{\bf Inaccurate face masks}: In previous datasets, the face masks are either rectangular, which may not completely cover the facial parts in the original video frame, or the convex hull of landmarks on eyebrow and lower lip, which leaves the boundaries of the mask visible. We improve the mask generation step for Celeb-DF. We first synthesize a face with more surrounding context, so as to completely cover the original facial parts after warping. We then create a smoothness mask based on the landmarks on eyebrow and interpolated points on cheeks and between lower lip and chin. The difference in mask generation used in existing datasets and Celeb-DF is highlighted in Fig.\ref{fig:mask} with an example.

\smallskip
\noindent{\bf Temporal flickering}: We reduce temporal flickering of synthetic faces in the DeepFake videos by incorporating temporal correlations among the detected face landmarks. Specifically, the temporal sequence of the face landmarks are filtered using a Kalman smoothing algorithm to reduce imprecise variations of landmarks in each frame.  

\subsection{Visual Quality}

The refinements to the synthesis algorithm improve the visual qualities of the DeepFake videos in the Celeb-DF dataset, as demonstrated in Fig.\ref{fig:demo}.
We would like have a more quantitative evaluation of the improvement in visual quality of the DeepFake videos in Celeb-DF and compare with the previous DeepFake datasets. Ideally, a reference-free face image quality metric is the best choice for this purpose. However, unfortunately, to date there is no such metric that is agreed upon and widely adopted. 

Instead, we follow the face in-painting work \cite{sun2018natural} and use the  Mask-SSIM score \cite{ma2017pose} as a referenced quantitative metric of visual quality of synthesized DeepFake video frames. Mask-SSIM corresponds to the SSIM score \cite{wang2004image} between the head regions (including face and hair) of the DeepFake video frame and the corresponding original video frame, \ie, the head region of the original target is the reference for visual quality evaluation. As such, low Mask-SSIM score may be due to inferior visual quality as well as changes of the identity from the target to the donor. On the other hand, since we only compare frames from DeepFake videos, the errors caused by identity changes are biased in a similar fashion to all compared datasets. Therefore, the numerical values of Mask-SSIM may not be meaningful to evaluate the absolute visual quality of the synthesized faces, but the difference between Mask-SSIM reflects the difference in visual quality.

\begin{table}[t]
\centering
 \resizebox{\linewidth}{!}{
  \begin{tabular}{|c|c|c|c|c|c|c|c|}
    \hline
    \multirow{2}{*}{Datasets}
    & \multirow{2}{*}{UADFV}
    & \multicolumn{2}{c|}{DF-TIMIT}
    & \multirow{2}{*}{FF-DF}
    & \multirow{2}{*}{DFD}
    & \multirow{2}{*}{DFDC} 
    & \multirow{2}{*}{\bf Celeb-DF} \\
    \cline{3-4}
    & & LQ & HQ & & &  & \\
    \hline
    \hline
    Mask & \multirow{2}{*}{\large 0.82} & \multirow{2}{*}{\large 0.80} & \multirow{2}{*}{\large 0.80} & \multirow{2}{*}{\large 0.81} & \multirow{2}{*}{\large 0.88} & \multirow{2}{*}{\large 0.84} & \multirow{2}{*}{\large {\bf 0.92}} \\ 
    -SSIM & & & & & & & \\
    \hline
  \end{tabular}
  }
  \caption{\em \small Average Mask-SSIM scores of different DeepFake datasets. {Computing Mask-SSIM requires exact corresponding pairs of DeepFake synthesized frames and original video frames, which is not the case for DFD and DFDC. For these two datasets, we calculate the Mask-SSIM on videos that we have exact correspondences, \ie, $311$ videos in DFD and $2,025$ videos in DFDC.}}
  \label{table:mask-ssim}
  \vspace{-1.5em}
\end{table}


The Mask-SSIM score takes value in the range of $[0,1]$ with higher value corresponding to better image quality. Table \ref{table:mask-ssim} shows the average Mask-SSIM scores for all compared datasets, with Celeb-DF having the highest scores. This confirms the visual observation that Celeb-DF has improved visual quality, as shown in Fig.\ref{fig:demo}.

\section{Evaluating DeepFake Detection Methods}

\begin{table*}[t]
\small
\centering
  \resizebox{\linewidth}{!}{
  \begin{tabular}{|c|c|c|c|c|}
    \hline
    Methods & Model Type & Training Dataset & Repositories & Release Date \\
    \hline
    \hline
    Two-stream \cite{zhou2017two} & GoogLeNet InceptionV3 \cite{googlenet} & SwapMe \cite{zhou2017two} & Unpublished code provided by the authors & 2018.03 \\
    \hline
    MesoNet \cite{afchar2018mesonet} & Designed CNN & Unpublished & \url{https://github.com/DariusAf/MesoNet} & 2018.09 \\
    \hline
    HeadPose \cite{yang2018exposing} & SVM & UADFV \cite{yang2018exposing} & \url{https://bitbucket.org/ericyang3721/headpose_forensic/} & 2018.11 \\
    \hline
    FWA \cite{li2019exposing} & ResNet-50 \cite{he2016deep} & Unpublished & \url{https://github.com/danmohaha/CVPRW2019_Face_Artifacts} & 2018.11 \\
    \hline
    VA-MLP \cite{matern2019exploiting} & Designed CNN & \multirow{2}{*}{Unpublished} & \multirow{2}{*}{\url{https://github.com/FalkoMatern/Exploiting-Visual-Artifacts}} & \multirow{2}{*}{2019.01} \\
    \cline{1-2}
    VA-LogReg \cite{matern2019exploiting} & Logistic Regression Model & & &  \\
    \hline
    Xception \cite{roessler2019faceforensics++} & XceptionNet \cite{chollet2017xception} & FaceForensics++ \cite{roessler2019faceforensics++} & \url{ https://github.com/ondyari/FaceForensics} & 2019.01 \\
    \hline
    Multi-task \cite{nguyen2019multi} & Designed CNN & FaceForensics \cite{rossler2018faceforensics} & \url{https://github.com/nii-yamagishilab/ClassNSeg} & 2019.06 \\
    \hline
    Capsule \cite{nguyen2019capsulev2} & Designed CapsuleNet \cite{sabour2017dynamic} & FaceForensics++ & \url{https://github.com/nii-yamagishilab/Capsule-Forensics-v2} & 2019.10 \\
    \hline
    DSP-FWA & SPPNet \cite{he2015spatial} & Unpublished & \url{https://github.com/danmohaha/DSP-FWA} & 2019.11 \\
    \hline
  \end{tabular}
  }
  \caption{\em \small Summary of compared DeepFake detection methods. See texts for more details.}
  \label{table:methods_stat}
   \vspace{-1.5em}
\end{table*}

Using Celeb-DF and other existing DeepFake datasets, we perform the most comprehensive performance evaluation of DeepFake detection to date, with the largest number of DeepFake detection methods and datasets considered. There are two purposes of this evaluation. First, using the average detection performance as an indicator of the challenge levels of various DeepFake datasets, we further compare Celeb-DF with existing DeepFake datasets. Furthermore, we survey the performance of the current DeepFake detection methods on a large diversity of DeepFake videos, in particular, the high-quality ones in Celeb-DF.

\subsection{Compared DeepFake Detection Methods}

We consider nine DeepFake detection methods in our experiments. {Because of the need to run each method on the Celeb-DF dataset, we choose only those that have code and the corresponding DNN-model publicly available or obtained from the authors directly.}
\begin{compactitem}
    \item {\bf Two-stream} \cite{zhou2017two} uses a two-stream CNN to achieve state-of-the-art performance in general-purpose image forgery detection. The underlying CNN is the GoogLeNet InceptionV3 model \cite{googlenet} trained on the SwapMe dataset \cite{zhou2017two}. We use it as a baseline to compare other dedicated DeepFake detection methods. 
    
    \item {\bf MesoNet} \cite{afchar2018mesonet} is a CNN-based DeepFake detection method targeting on the mesoscopic properties of images. The model is trained on unpublished DeepFake datasets collected by the authors. We evaluate two variants of MesoNet, namely,  {\em Meso4} and {\em MesoInception4}. {Meso4 uses conventional convolutional layers, while MesoInception4 is based on the more sophisticated Inception modules \cite{szegedy2015going}.} 
    
    \item {\bf HeadPose} \cite{yang2018exposing} detects DeepFake videos using the inconsistencies in the head poses of the synthesized videos, based on a SVM model on estimated 3D head orientations from each video. The SVM model in this method is trained on the UADFV dataset.    
    
    \item {\bf FWA} \cite{li2019exposing} detects DeepFake videos using a ResNet-50 \cite{he2016deep} to expose the face warping artifacts introduced by the resizing and interpolation operations in the basic DeepFake maker algorithm. This model is trained on self-collected face images.

    \item {\bf VA} \cite{matern2019exploiting} is a recent DeepFake detection method based on capturing visual artifacts in the eyes, teeth and facial contours of the synthesized faces. There are two variants of this method: VA-MLP is based on a multilayer feedforward neural network classifier, and VA-LogReg uses a simpler logistic regression model. These models are trained on unpublished dataset, of which real images are cropped from CelebA dataset \cite{liu2015deep} and the DeepFake videos are from {\tt YouTube}.   
    
    \item {\bf Xception} \cite{roessler2019faceforensics++} corresponds to a DeepFake detection method based on the XceptionNet model \cite{chollet2017xception} trained on the FaceForensics++ dataset. There are three variants of Xception, namely, {{\em Xception-raw}, {\em Xception-c23} and {\em Xception-c40}: {\em Xception-raw} are trained on raw videos, while {\em Xception-c23} and {\em Xception-c40} are trained on H.264 videos with medium (23) and high degrees (40) of compression, respectively.} 
    
     \item {\bf Multi-task} \cite{nguyen2019multi} is another recent DeepFake detection method that uses a CNN model to simultaneously detect manipulated images and segment manipulated areas as a multi-task learning problem. This model is trained on the FaceForensics dataset \cite{rossler2018faceforensics}.
    
    \item {\bf Capsule} \cite{nguyen2019capsulev2} uses capsule structures \cite{sabour2017dynamic} based on a VGG19 \cite{simonyan2014very} network as the backbone architecture for DeepFake classification. This model is trained on the FaceForensics++ dataset.
    
    \item {\bf DSP-FWA} is a recently further improved method based on FWA, which includes a spatial pyramid pooling (SPP) module \cite{he2015spatial} to better handle the variations in the resolutions of the original target faces. This method is trained on self-collected face images.
\end{compactitem}
A concise summary of the underlying model, source code, and training datasets of the DeepFake detection methods considered in our experiments is given in Table \ref{table:methods_stat}.

\subsection{Experimental Settings}

We evaluate the overall detection performance using the area under ROC curve (AUC) score at the frame level for all key frames. There are several reasons for this choice. First, all compared methods analyze individual frames (usually key frames of a video) and output a classification score for each frame. Using frame-level AUC thus avoids differences caused by different approaches to aggregating frame-level scores for each video. Second, using frame level AUC score obviates the necessity of calibrating the classification outputs of these methods across different datasets. To increase robustness to numerical imprecision, the classification scores are rounded to five digits after the decimal point, \ie, with a precision of $10^{-5}$. As the videos are compressed, we perform evaluations only on the key frames. 

we compare performance of each detection method using the inference code and the published pre-trained models. This is because most of these methods do not have published code for training the machine learning models. As such, we could not practically re-train these models on all datasets we considered. We use the default parameters provided with each compared detection method. 

\subsection{Results and Analysis}  

\begin{table*}[t]
\small
\centering
  \begin{tabular}{|c|c|p{1.3cm}<{\centering}|p{1.3cm}<{\centering}|c|c|c|c|}
    \hline
    \multirow{2}{*}{Methods$\downarrow$ Datasets$\rightarrow$}
    & \multirow{2}{*}{UADFV \cite{yang2018exposing}}
    & \multicolumn{2}{c|}{DF-TIMIT \cite{korshunov2018deepfakes}}
    & \multirow{2}{*}{FF-DF \cite{roessler2019faceforensics++} }
    & \multirow{2}{*}{DFD \cite{DDD_GoogleJigSaw2019}}
    & \multirow{2}{*}{DFDC \cite{dolhansky2019deepfake}} 
    & \multirow{2}{*}{\bf Celeb-DF} \\
    \cline{3-4}
    & & LQ & HQ & & &  & \\
    \hline
    \hline
    Two-stream \cite{zhou2017two}   
    & 85.1 & 83.5 & 73.5 & 70.1 & 52.8 & 61.4 & 53.8 \\
    \hline
    Meso4 \cite{afchar2018mesonet}
    & 84.3 & 87.8 & 68.4 & 84.7 & 76.0 & 75.3 & 54.8 \\
    MesoInception4 
    & 82.1 & 80.4 & 62.7 & 83.0 & 75.9 & 73.2 & 53.6 \\
    \hline
    HeadPose \cite{yang2018exposing}
    & 89.0 & 55.1 & 53.2 & 47.3 & 56.1 & 55.9 & 54.6 \\
    \hline
    FWA \cite{li2019exposing}
    & 97.4 & 99.9 & 93.2 & 80.1 & 74.3 & 72.7 & 56.9 \\
    \hline
    VA-MLP \cite{matern2019exploiting}
    & 70.2 & 61.4 & 62.1 & 66.4 & 69.1 & 61.9 & 55.0 \\
    VA-LogReg 
    & 54.0 & 77.0 & 77.3 & 78.0 & 77.2 & 66.2 & 55.1 \\
    \hline
    Xception-raw \cite{roessler2019faceforensics++} 
    & 80.4 & 56.7 & 54.0 & {\bf 99.7} & 53.9 & 49.9 & 48.2 \\
    Xception-c23 
    & 91.2 & 95.9 & 94.4 & 99.7 & {\bf 85.9} & 72.2 & 65.3 \\
    Xception-c40 
    & 83.6 & 75.8 & 70.5 & 95.5 & 65.8 & 69.7 & {\bf 65.5} \\
    \hline
    Multi-task \cite{nguyen2019multi}
    & 65.8 & 62.2 & 55.3 & 76.3 & 54.1 & 53.6 & 54.3 \\
    \hline
    Capsule \cite{nguyen2019capsulev2} 
    & 61.3 & 78.4 & 74.4 & 96.6 & 64.0 & 53.3 & 57.5 \\
    \hline
    DSP-FWA
    & {\bf 97.7} & {\bf 99.9} & {\bf 99.7} & 93.0 & 81.1 & {\bf 75.5} & 64.6 \\
    \hline
  \end{tabular}
  \caption{\em \small Frame-level AUC scores ($\%$) of various methods on compared datasets. Bold faces correspond to the top performance.}
  \label{table:compare_det}
  \vspace{-.5em}
\end{table*}

\begin{figure}[t]
\centering
\includegraphics[width=0.9\linewidth]{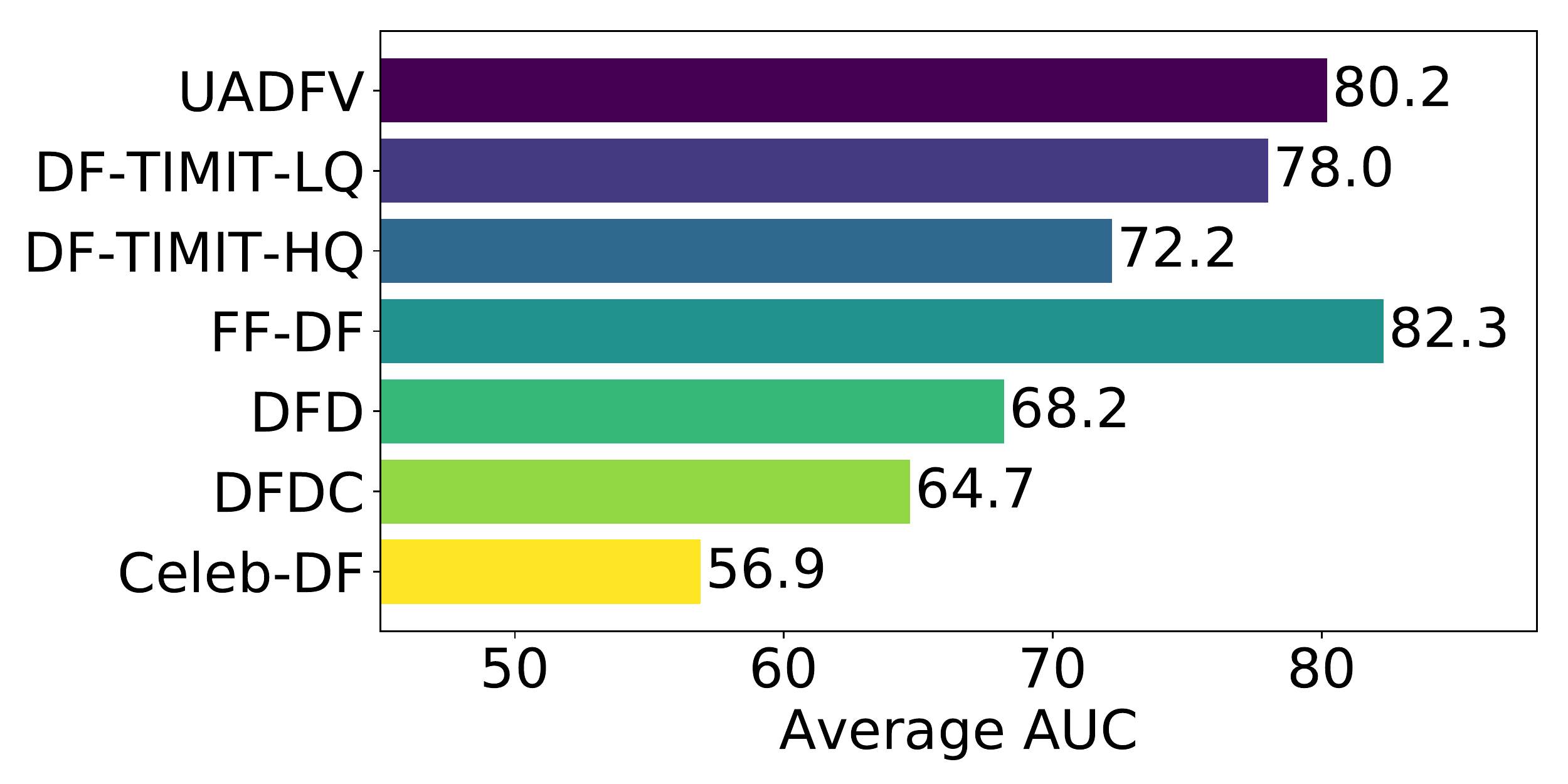}
\vspace{-0.2cm}
\caption{\small \em Average AUC performance of all detection methods on each dataset.}
\label{fig:dataset_auc}
\vspace{-1.em}
\end{figure}

\begin{figure*}[t]
    \hspace{0.5cm}
   \includegraphics[width=.3\textwidth]{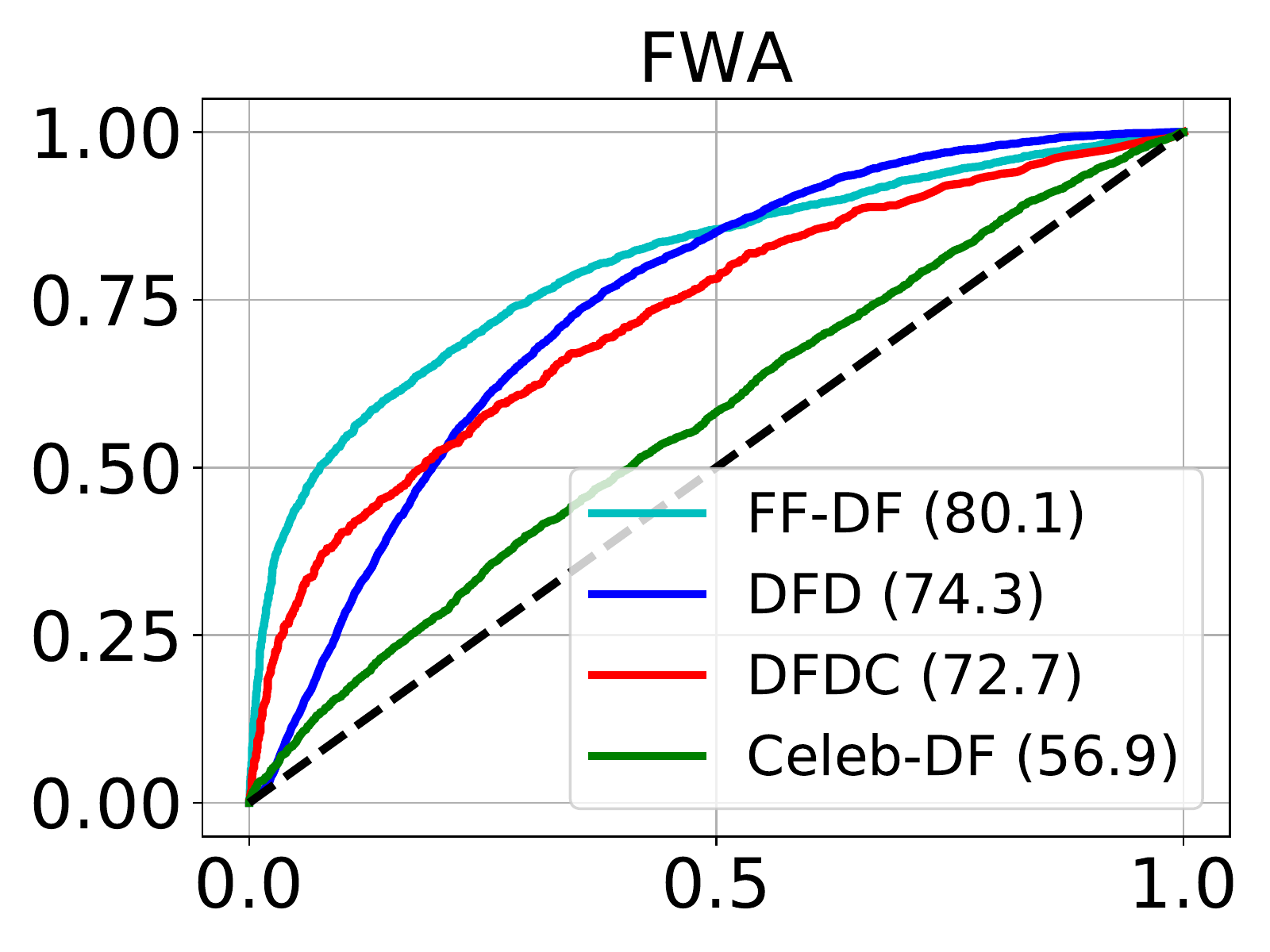}
   \includegraphics[width=.3\textwidth]{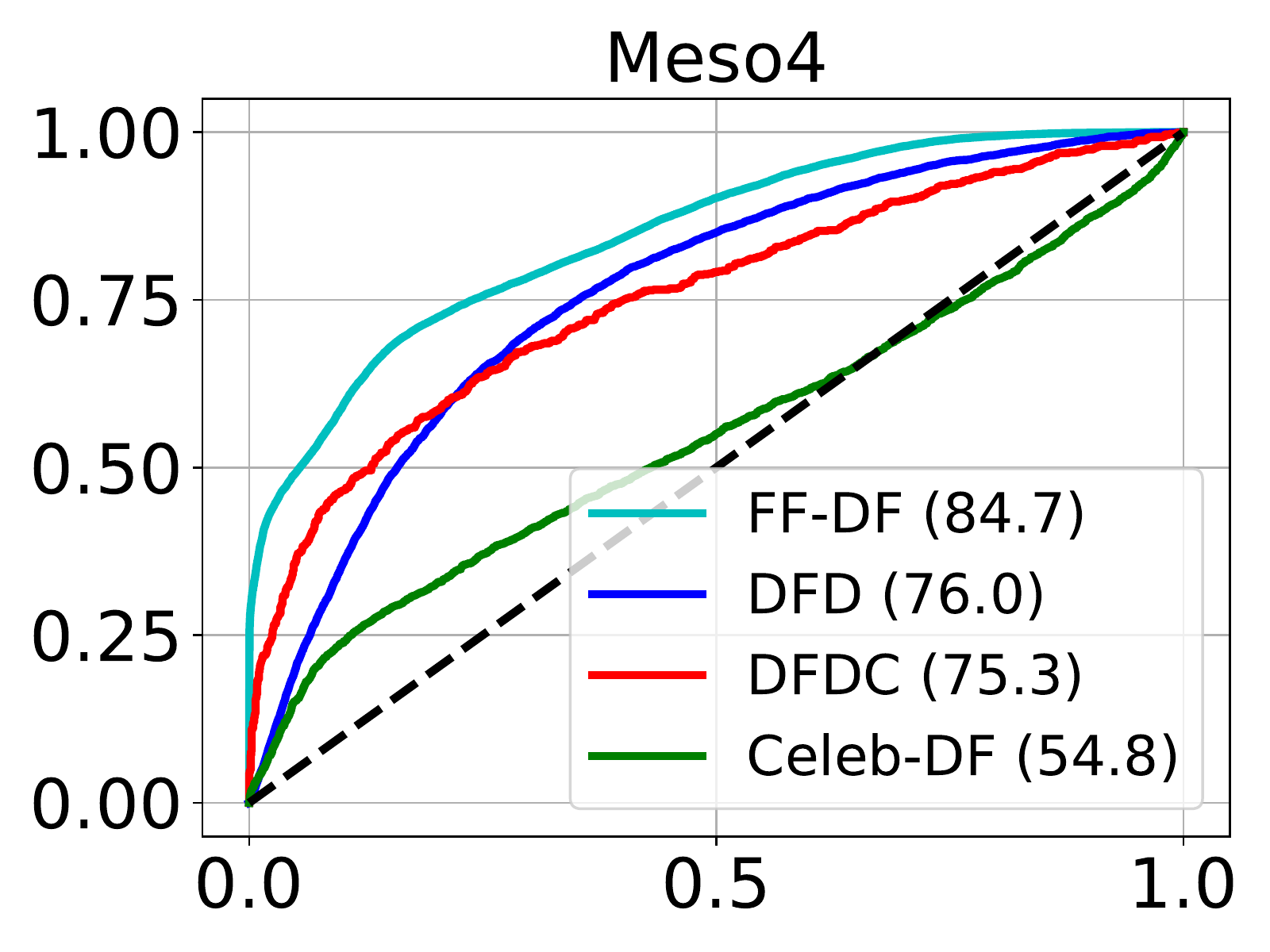}
   \includegraphics[width=.3\textwidth]{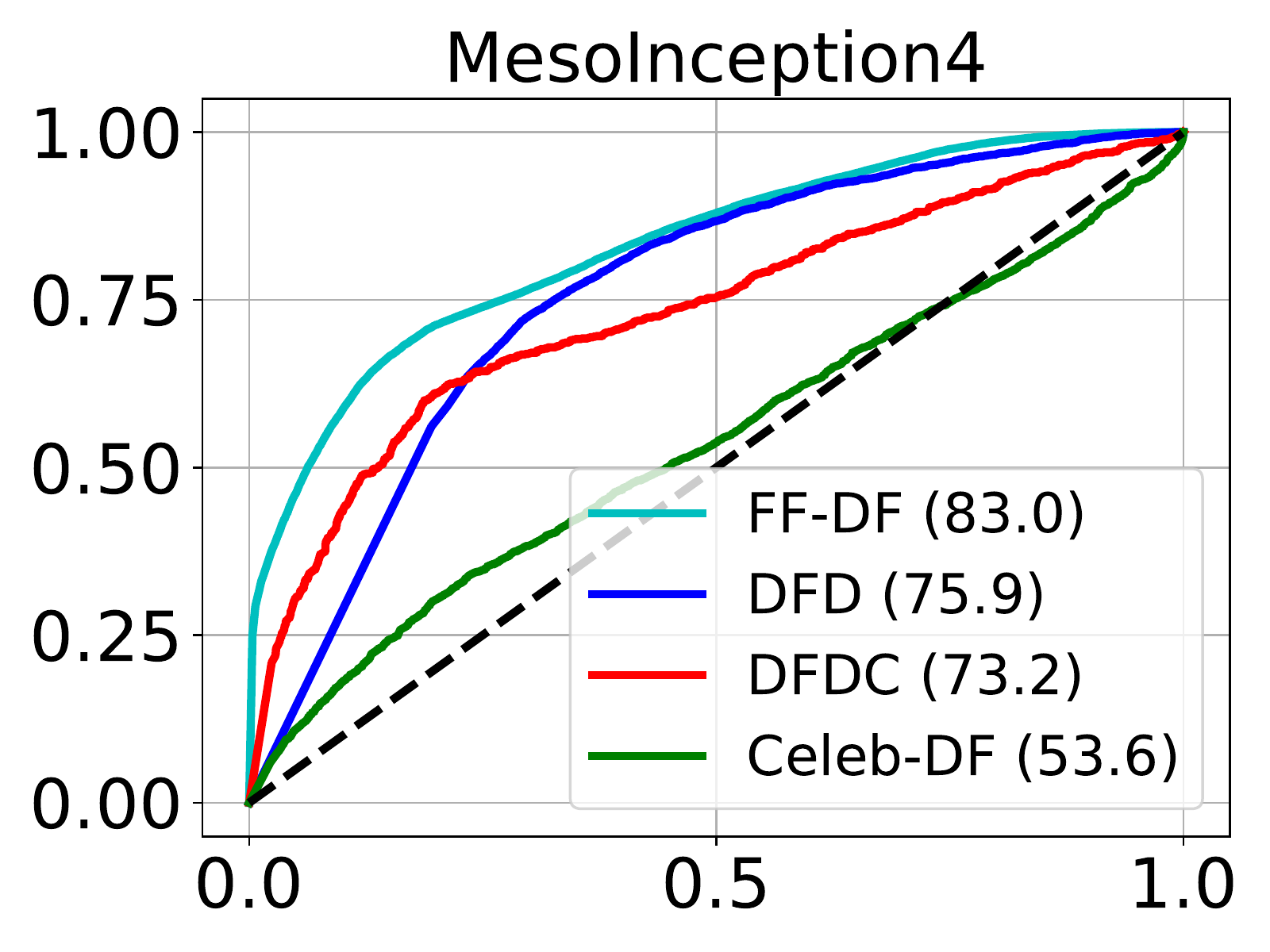} 
   
  \hspace{0.5cm} 
  \includegraphics[width=.3\textwidth]{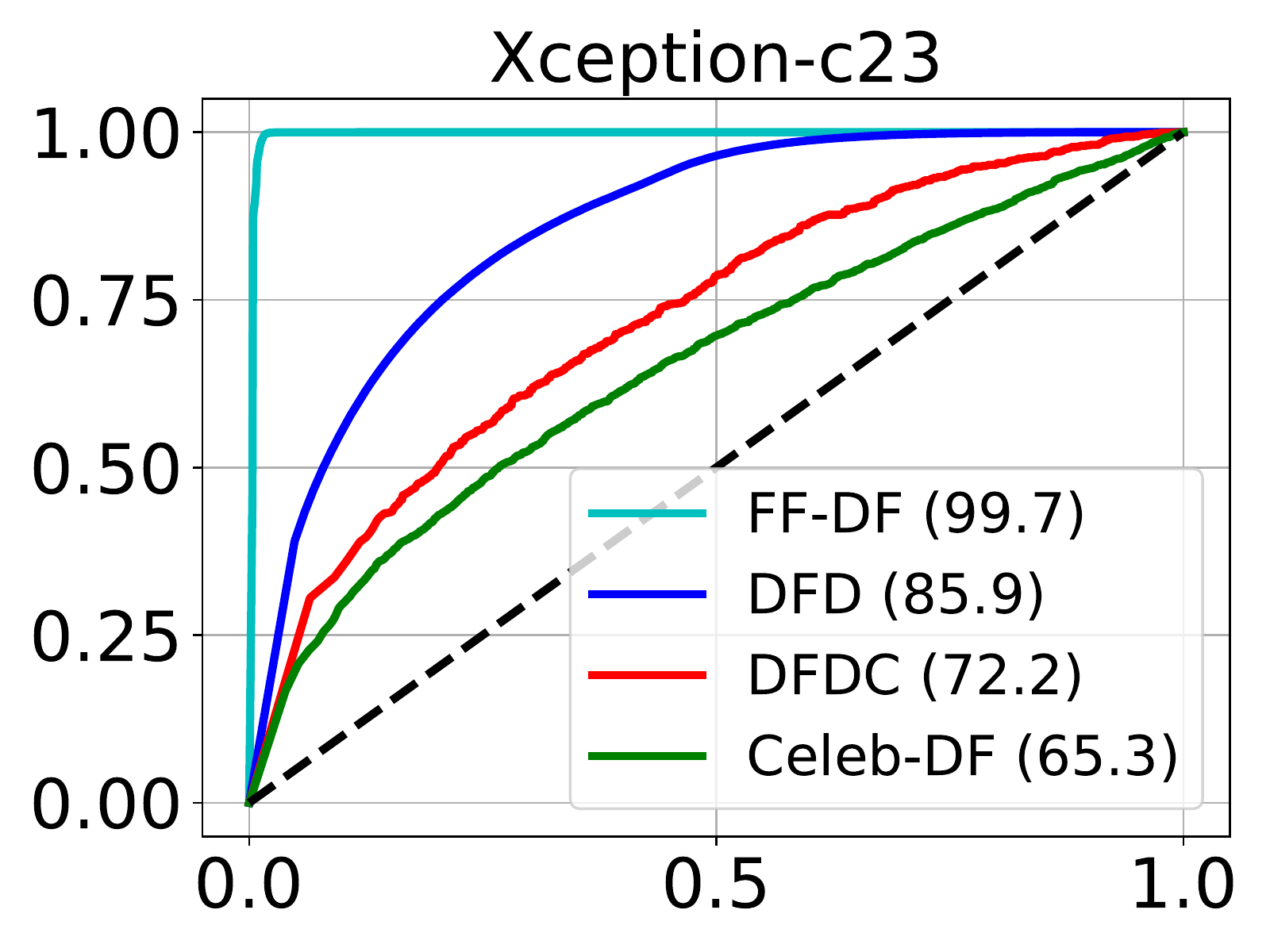} 
   \includegraphics[width=.3\textwidth]{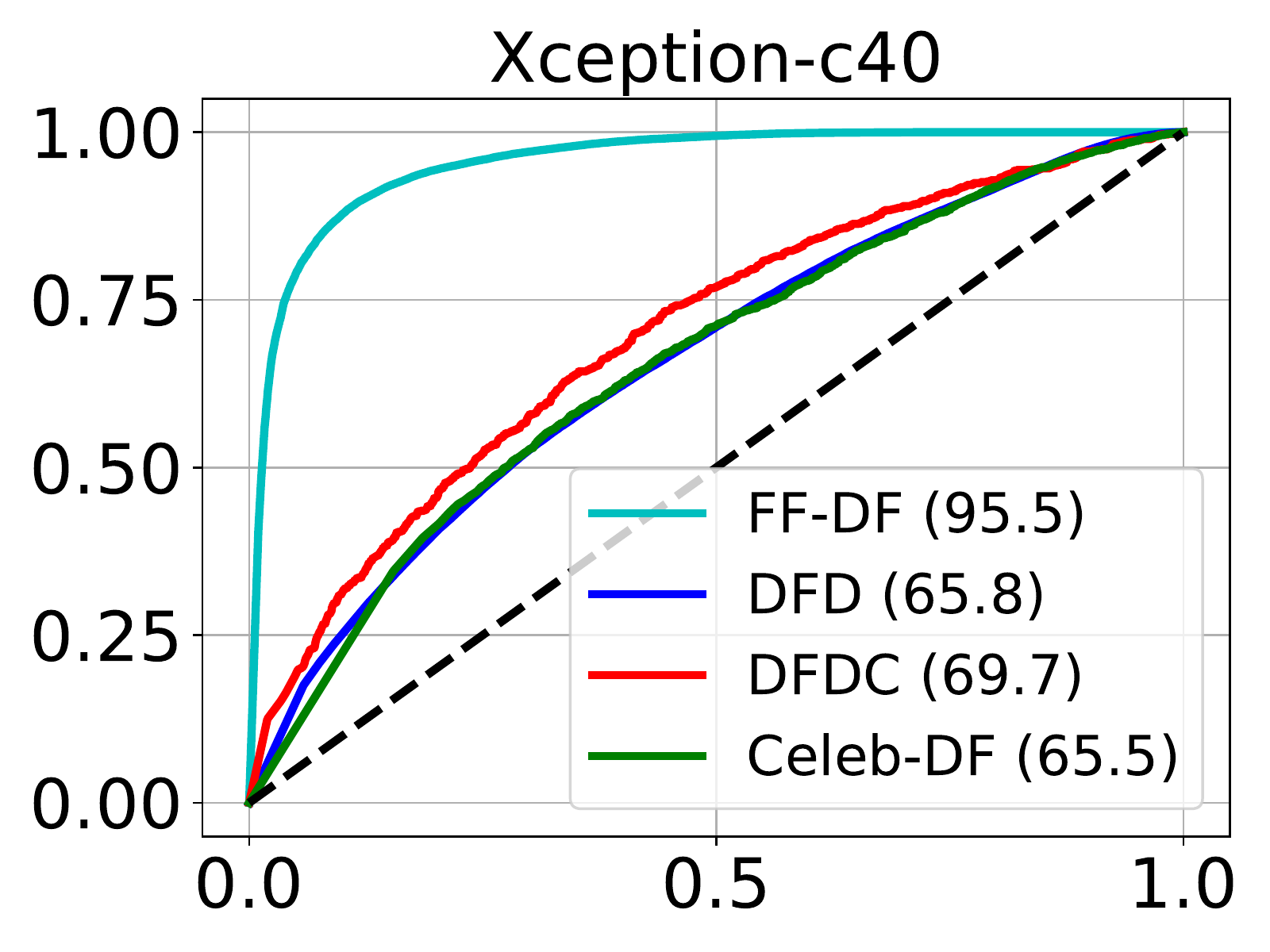}
   \includegraphics[width=.3\textwidth]{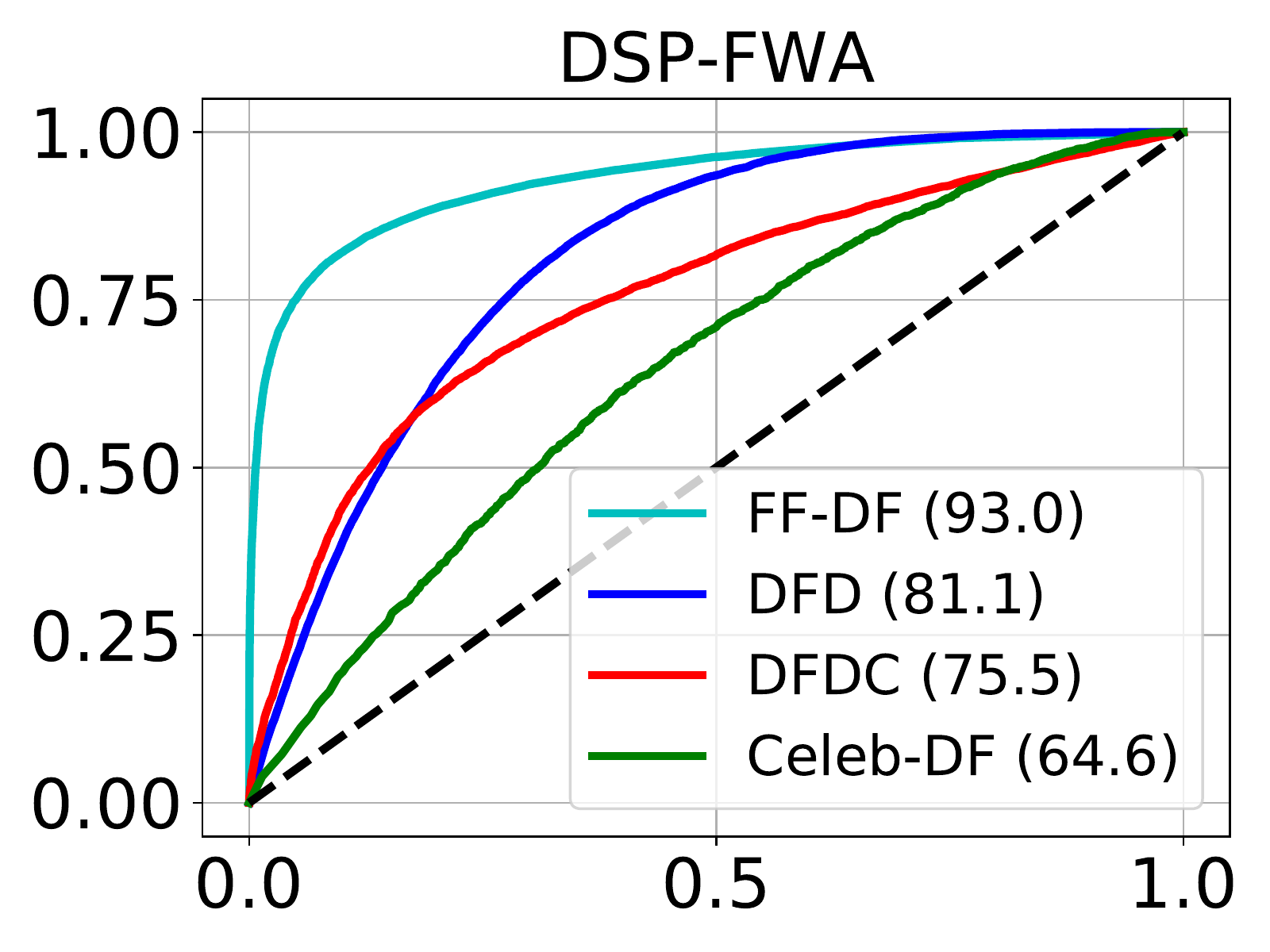}  
    \vspace{-0.6em}
    \caption{\em \small ROC curves of six state-of-the-art detection methods (FWA, Meso4, MesoInception4, Xception-c23, Xception-40 and DSP-FWA) on four largest datasets (FF-DF, DFD, DFDC and Celeb-DF).}
    \label{fig:auc_plots}
    \vspace{-1em}
\end{figure*}

In Table \ref{table:compare_det} we list individual frame-level AUC scores of all compared DeepFake detection methods over all datasets including Celeb-DF, and Fig.\ref{fig:auc_plots} shows the frame-level ROC curves of several top detection methods on several datasets. 

Comparing different datasets, in Fig.\ref{fig:dataset_auc}, we show the average frame-level AUC scores of all compared detection methods on each dataset. Celeb-DF is in general the most challenging to the current detection methods, and their overall performance on Celeb-DF is lowest across all datasets. These results are consistent with the differences in visual quality. Note many current detection methods predicate on visual artifacts such as low resolution and color mismatch, which are improved in synthesis algorithm for the Celeb-DF dataset. Furthermore, the difficulty level for detection is clearly higher for the second generation datasets (DFD, DFDC, and Celeb-DF, with average AUC scores lower than $70\%$), while some detection methods achieve near perfect detection on the first generation datasets (UADFV, DF-TIMIT, and FF-DF, with average AUC scores around $80\%$). 
\begin{figure}[t]
\centering
\includegraphics[width=\linewidth]{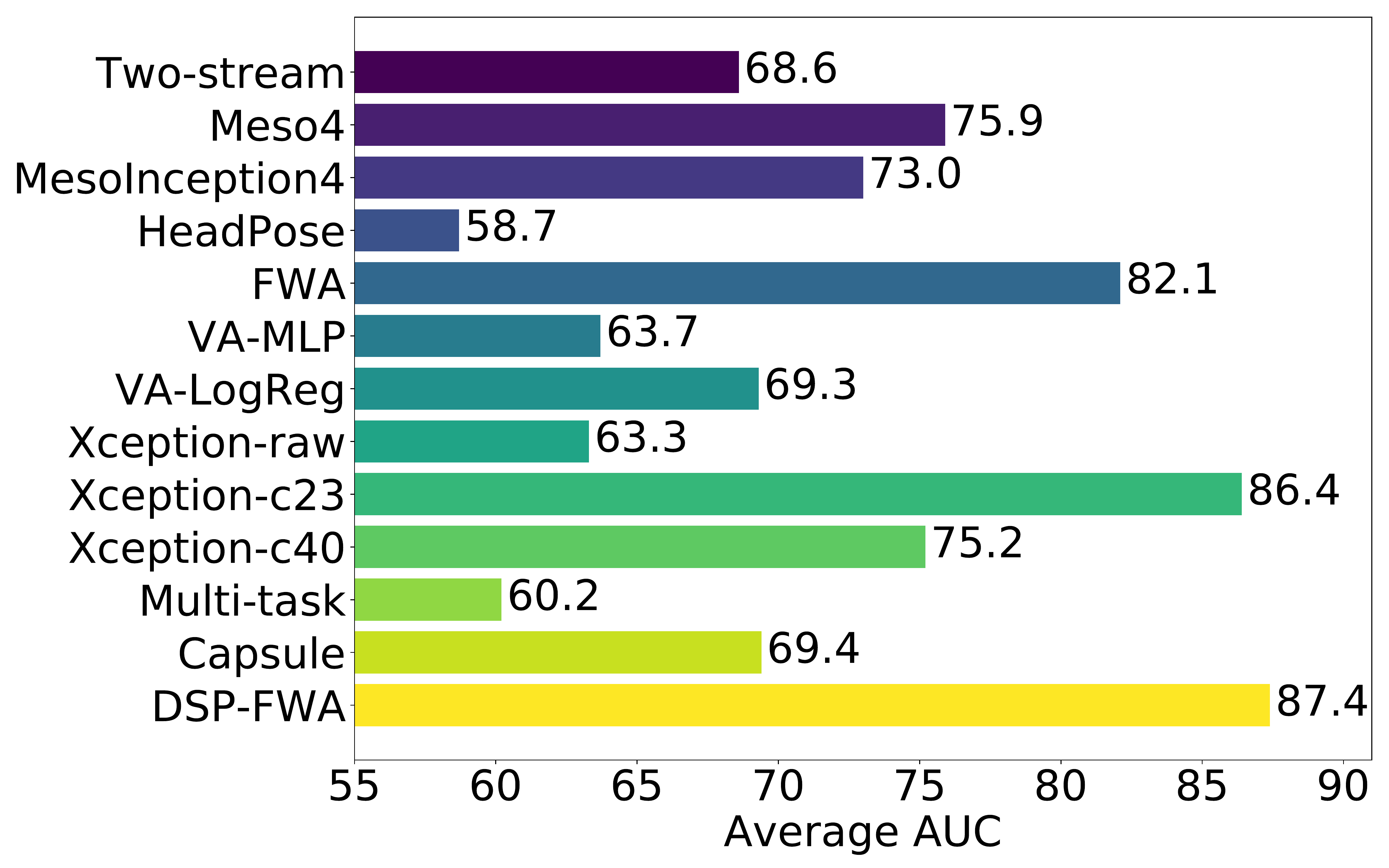}
\vspace{-0.6cm}
\caption{\small \em Average AUC performance of each detection method on all evaluated datasets.}
\label{fig:detectors_auc}
\vspace{-2.em}
\end{figure}

\begin{table}[t]
\small
\centering
  \begin{tabular}{|c|c|c|c|}
    \hline
     & Original & c23 & c40 \\
    \hline
    \hline
    FWA   & 56.9 & 54.6 & 52.2 \\
    \hline
    Xception-c23  & 65.3 & 65.5 & 52.5 \\
    \hline
    Xception-c40  & 65.5 & 65.4 & 59.4 \\
    \hline
    DSP-FWA   & 64.6 & 57.7 & 47.2 \\
    \hline
  \end{tabular}
  \caption{\em \small AUC performance of four top detection methods on original,  medium (23) and high (40) degrees of H.264 compressed Celeb-DF respectively.}
  \label{table:compression}
  \vspace{-1.7em}
\end{table}

In term of individual detection methods, Fig.\ref{fig:detectors_auc} shows the comparison of average AUC score of each detection method on all DeepFake datasets. These results show that detection has also made progress with the most recent DSP-FWA method achieves the overall top performance ($87.4\%$).

As online videos are usually recompressed to different formats (MPEG4.0 and H264) and in different qualities during the process of uploading and redistribution, it is also important to evaluate the robustness of detection performance with regards to video compression. {Table \ref{table:compression} shows the average frame-level AUC scores of four state-of-the-art DeepFake detection methods on original MPEG4.0 videos, and medium (23), and high (40) degrees of H.264 compressed videos of Celeb-DF, respectively. The results show that the performance of each method is reduced along with the compression degree increased. In particular, the performance of FWA and DSP-FWA degrades significantly on recompressed video, while the performance of Xception-c23 and Xception-c40 is not significantly affected. This is expected because the latter methods were trained on compressed H.264 videos such that they are more robust in this setting.}

\vspace{-0.2cm}
\section{Conclusion}
\vspace{-0.1cm}
We present a new challenging large-scale dataset for the development and evaluation of DeepFake detection methods. The Celeb-DF dataset reduces the gap in visual quality of DeepFake datasets and the actual DeepFake videos circulated online. Based on the Celeb-DF dataset, we perform a comprehensive performance evaluation of current DeepFake detection methods, and show that there is still much room for improvement. 

For future works, the foremost task is to enlarge the Celeb-DF dataset and improve the visual quality of the synthesized videos. This entails improving the running efficiency and model structure of the current synthesis algorithm. Furthermore, while the forgers can improve the visual quality in general, they may also adopt {\em anti-forensic} techniques, which aim to hide traces of DeepFake synthesis on which the detection methods predicate. Anticipating such counter-measures at the forgers' disposal, we aim to incorporate anti-forensic techniques in Celeb-DF. 

\smallskip
\noindent {\bf Acknowledgement}. {\small This material is based upon work supported by NSF under Grant No (IIS-1816227).  Any opinions, findings, and conclusions or recommendations expressed in this material are those of the author(s) and do not necessarily reflect the views of NSF.}
 
\newpage
{\small
\bibliographystyle{ieee_fullname}
\bibliography{egbib}
}

\end{document}